\documentclass[twocolumn,aps,pra,superscriptaddress,nofootinbib,a4paper,showpacs]{revtex4-1}
\usepackage{amsmath}
\usepackage{graphicx}

\newcommand{\be}{\begin{eqnarray}}
\newcommand{\ee}{\end{eqnarray}}
\newcommand{\beq}{\begin{eqnarray}}
\newcommand{\enq}{\end{eqnarray}}
\newcommand{\BEQ}{\begin{eqnarray}}
\newcommand{\ENQ}{\end{eqnarray}}
\newcommand{\EEQ}{\end{eqnarray}}
\newcommand{\forget}[1]{}
\newcommand{\nn}{\nonumber}

\begin{document}
\bibliographystyle{apsrmp}
\title{Not throwing out the baby with the bathwater: Bell's condition of local causality mathematically `sharp and clean'}
%claryfing requirements of sufficiency and completeness in Bell-type analysis}
%\title{Claryfing requirements of sufficiency and completeness in Bell-type analysis. Bell on ''sufficiently well specified beables''}
%\footnote{Dedicated to Hanneke Jansen (19??-2008).}}
\author{M.P. Seevinck}
\email{m.p.seevinck@science.ru.nl}
\affiliation{Institute for Mathematics, Astrophysics and Particle Physics, Faculty of Science, Radboud University Nijmegen, the Netherlands;}
\affiliation{Centre for the History of Philosophy and Science, Faculty of Philosophy, Radboud University Nijmegen, the Netherlands}
\affiliation{Institute for History and Foundations of Science, Utrecht University, the Netherlands}
\author{J. Uffink}
\affiliation{Institute for History and Foundations of Science, Utrecht University, the Netherlands}

\begin{abstract}\noindent
The  starting point of the present paper is Bell's notion of local causality and his own  sharpening of it so as to provide for mathematical formalisation. Starting with Norsen's (2007 \cite{norsen07}, 2009 \cite{norsen09}) analysis of this formalisation, it is subjected to a critique that reveals two crucial aspects that have so far not been properly taken into account.  These are (i) the correct understanding of the notions of sufficiency, completeness and redundancy involved;  and (ii) the fact that the apparatus settings and measurement  outcomes have very different theoretical roles in the candidate theories under study. Both aspects are not adequately incorporated in the standard formalisation, and we will therefore do so. The upshot of our analysis is a more detailed, sharp and clean mathematical expression of the condition of local causality.  A preliminary analysis of the repercussions of our proposal shows that it is able to locate exactly where and how the notions of locality and causality are involved in formalising Bell's condition of local causality.
 \end{abstract}


\maketitle
%~\vskip0.00001cm
\begin{quote}~\vskip0.01cm\emph{Now it is precisely in cleaning up 
intuitive ideas for mathematics that one is likely to throw out the 
baby with the bathwater.}\vskip0.5em $~~~~~~~~~~~~~~~~~~~~~~~~~$J.S. Bell (1990) \citep[p.~106]{bell90}
\end{quote}
~\vskip0.5cm
%\tableofcontents 
%
%
\section{Introduction}\noindent
Despite the existence of many mathematically precise results concerning Bell's theorem, there continues to be controversy over just what the ingredients of the theorem are, and what the theorem together with the experiments confirming a violation of a Bell-inequality rules out. This is especially so with regard the locality and causality requirements involved.  Although recently great progress has been achieved in clarifying Bell's theorem and the crucial notions and assumptions involved\footnote{For example, see Cavalcanti (2008, \cite{cavalcanti}), Norsen (2007 \cite{norsen07}, 2009 \cite{norsen09}), Seevinck (2008 \cite{thesis}).}, we still believe there to be a crucial gap left open. This paper, which is still work in progress,  tries to fill in this gap.

In section \ref{intuition} we will present the intuitive notion of local causality and the way Bell himself further sharpened it so as to allow for mathematical formalisation. 
In section \ref{cleaning_up} Bell's formalisation of local causality will be commented on using the illuminating papers by Norsen (2007, \cite{norsen07}, 2009 \cite{norsen09}). We will closely follow Norsen's analysis\footnote{Parts of the present paper merely rehearse aspects of Norsen's analysis. See foonote \ref{worthwhile} for the reason of doing so.} but in doing so we will unearth two novel and crucial aspects that have not yet been adequately incorporated in the standard formalisation.  Firstly, section \ref{prelim_suff} indicates the intricate relationship of the notions of sufficiency, completeness and redundancy involved. Secondly, the very different theoretical role of settings and outcomes in the candidate theories under study is argued for in section \ref{diff_sett_out}. The latter can be rather easily performed but the first needs a rather extensive discussion. 
This is performed in section \ref{formalsuff}, and  which gives a mathematical account of the two different notions of sufficiency that are in play. It will use an important source of inspiration that has been overlooked in the debate so far. Namely, it will be argued that the concept of sufficiency, first formulated by R.A. Fisher (1922 \cite{fisher22}) in the context of mathematical statistics throws a relevant light on this debate. 

Next, the tools obtained in the course of our analysis are used in section \ref{babysection} to finally give a mathematically sharp and clean formulation of 
Bell's notion of local causality.
In section \ref{notout} the novel mathematical formalisation is analysed and it is precisely indicated where and how the notions of locality and causality are involved. 
Section \ref{envoi} marks the end of this paper by indicating that, indeed, the present paper is still work in progress, as the repercussions
of our mathematical formalisation of Bell's notion of local causality need still to be fully charted and confronted to those of other similar analysis% interpretations and motivations of Bell's notion of local causality
\footnote{This we hope to do in the near future (Seevink \& Uffink, 2010, \cite{inprogress}).}.

\section{The intuitive idea: Bell's local causality}\label{intuition}
\noindent
In the section entitled `Principle of local causality' of the very last article Bell wrote on the foundations of quantum theory  (published in 1990 and entitled 'La Nouvelle Cuisine' \cite{bell90}), Bell begins his explanation of the principle of local causality as follows:\footnote{Here we will mainly focus on Bell's formulation of this principle as presented in 'La Nouvelle Cuisine', Bell (1990) \cite{bell90}. This presentation we take to be the most definite and precise one Bell ever presented; it is overall consistent with earlier formulations Bell used to indicate this principle. See Norsen (2007) \cite{norsen07} for further elaboration and support of this claim.}
\begin{quote}
``The direct causes (and effects) of events are near by, and even the indirect causes (and effects) are no further away than permitted by the velocity of light.''  Bell (1990) \cite[p.~105]{bell90}
\end{quote}
\begin{figure}[h]
\includegraphics[scale=0.6]{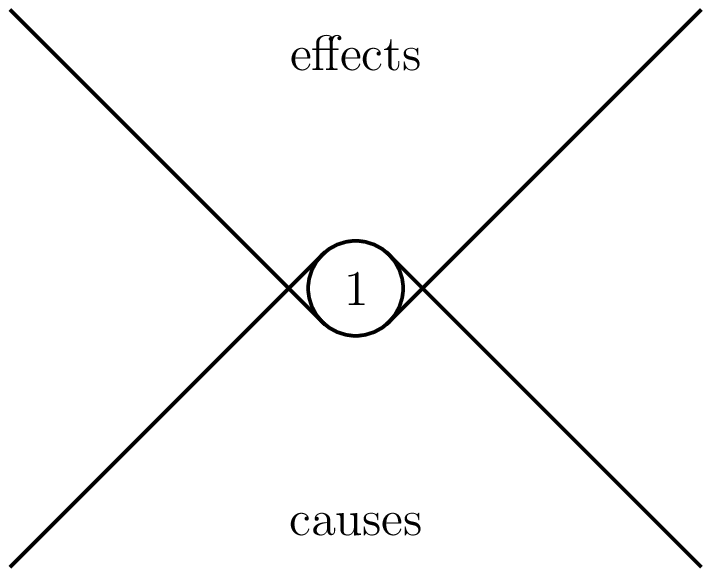}
\caption{``Space-time location of causes and effects of events in region 1.'' Figure (slightly modified) and caption taken from Bell (1990 \cite[p.~105]{bell90}).
} 
\label{figcausal}
\end{figure}
This locates the causes operating in a certain region in space-time in the backward light cone of that region and effects of anything occuring in that region in its forward light cone. See Fig.~\ref{figcausal}. But Bell remarks, the ``[t]he above principle is not yet sufficiently sharp and clean for mathematics''. He then continues (see Fig. \ref{figBell}):
\begin{figure}[h]
\includegraphics[scale=0.55]{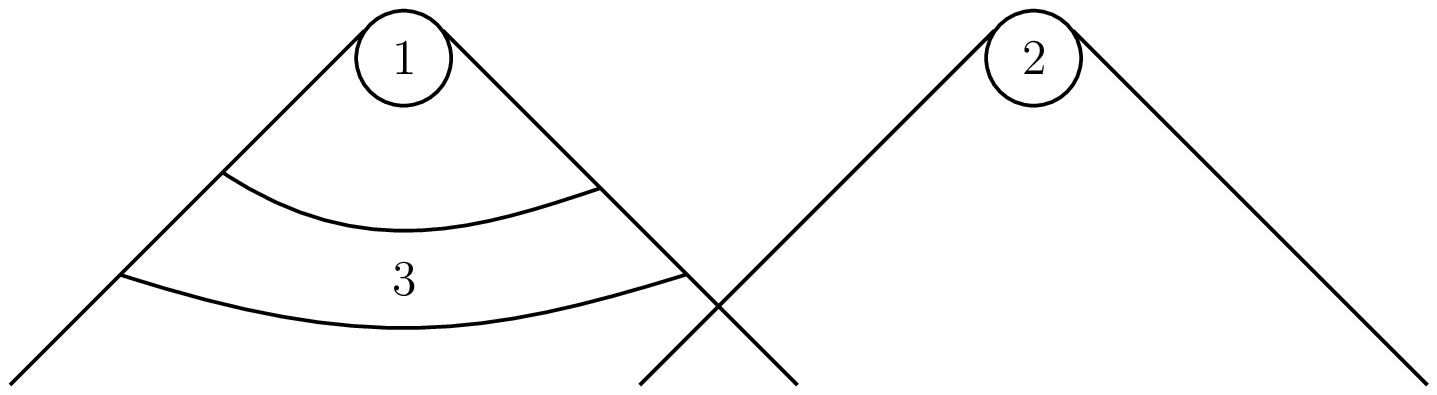}
\caption{``Full specification of what happens in 3 makes events in 2 irrelevant for predictions about 1 in a locally causal theory.'' Figure and caption taken from Bell (1990) \cite[p.~105]{bell90}.} 
\label{figBell}
\end{figure}
\begin{quote}
``A theory is said to be locally causal if the probabilities attached to values of local beables in a space-time region 1 are unaltered by a specification of values of local beables in a space-like separated region 2 when what happens in the backward light cone is already sufficiently specified, for example by a full specification of local beables in a spacetime region 3. It is important that region 3 completely shields off from 1 the overlap of the backward light cones of 1 and 2. And it is important th at events 3 be specified completely. Otherwise the traces in region 2 of causes of events in 1 could well supplement whatever else was being used for calculating probabilities about 1. The hypothesis is that any such information about 2 becomes redundant when 3 is specified completely.'' Bell (1990) \cite[p.~106]{bell90}
\end{quote}
Although this formulation is considerably sharper, it is not yet cleanly formulated in terms of mathematics. Probably for this reason Bell introduces some further notation and terminology in a subsequent discussion. % where he derives factorisability from his principle of local causality. 
He in effect introduces the space-time diagram of Fig. \ref{figBell_fact} that is adapted\footnote{See footnote \ref{norsen_fout}.} from Norsen's (2009) \cite{norsen09} highly illuminating paper.
\begin{figure}[h]
\includegraphics[scale=0.6]{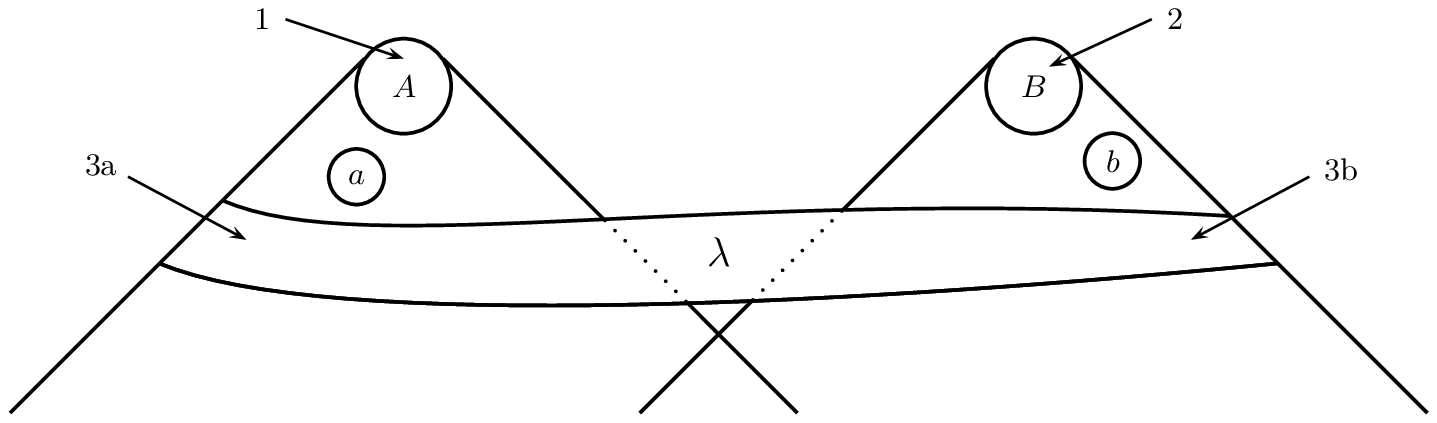}
\caption{Space-time diagram of the setup Bell considers. 
%Two experimenters, often called Alice and Bob, each measure an observable $a$ and $b$, respectively, so as to %obtain outcomes $A$ and $B$. SAY MORE. 
For explanation, see text. Figure adapted from Norsen (2009) \cite{norsen09}.
} 
\label{figBell_fact}
\end{figure}

This diagram encodes the setup Bell considers. It involves measurement on a bi-partite system (e.g., two particles emitted by a source) where each part is measured by a different party, called Alice and Bob respectively.
The outcomes of measurement are represented by beables $A$ (in region 1) and $B$ (in region 2) and the settings chosen by experimenters Alice and Bob are denoted by beables $a$ and $b$ respectively. The symbol $\lambda$ indicates the specification of the state of the bipartite system under study together with other relevant beables in the spacetime regions 3a and 3b.
%Here it is assumed that the beables and the settings are not dependent on each other, i.e., the `the variables $a$ and %$b$ can be considered to  \emph{free}\footnote{Bell also calls them `free variables', e.g. `at the whim of %experimenters'. (Bell, 1976) \cite{bell76}}, or \emph{random}' (Bell, 1990) \cite[p.~109]{bell90}: $\rho(\lambda|A,B)=\rho%(\lambda)$, with $\rho$ some density. 

The logic is now as follows. Consider a candidate theory that attempts to describe any correlations found between outcomes $A$ and $B$. Suppose region 3a shields off region 1 from the overlap of the past light cones of 1 and 2, and, likewise, that region 3b shields off region 2 from the overlap of the past light cones of 1 and 2 (see Fig. \ref{figBell_fact}). 
It is assumed that (in this candidate theory under study) $\lambda$ constitutes a complete specification\footnote{\label{norsen_fout}Norsen (2007 \cite{norsen07}, 2009 \cite{norsen09}) requires that such a complete specification of region 3a and 3b not only includes $\lambda$ but also the setting $a$ in region 3a and setting $b$ in region 3b, respectively.  For this purpose he has appropriately located the spacetime location of settings to overlap with region 3 in Fig. \ref{figBell_fact}.
We believe, however, that this need not be done, and in fact should not be done. In remark (two) on page \pageref{(two)} this is further explained.} of the beables in region 3a and 3b.
%
%It is assumed that (in this candidate theory under study) $\lambda$ and $a$ jointly constitute a complete specification of the %beables in region 3a. In the same fashion, $\lambda$ and $b$ are assumed to jointly constitute a complete specification of the %beables in region 3b.

With all this implicitly in place, Bell continues and applies his principle of local causality to this setup:  
\begin{quote}
``Invoking local causality, and the assumed completeness of \ldots $\lambda$ \ldots we declare redundant certain of the conditional variables in the last expression because they are at space-like separation from the result in question."
Bell (1990)\cite[p.~109]{bell90}\end{quote} 
Thus the  specification of  $\lambda$ makes both $B$ and $b$ redundant for prediction about $A$, and both $A$ and $a$ redundant for prediction about $B$.

This finally allows for a clean formulation in mathematics of the principle. For now we follow Norsen (2007) \cite{norsen07} in claiming that this indeed gives (but, see our critique later on)
\begin{align}\label{LCmath}
\begin{array}{l}
P(A|a,b,B,\lambda)=P(A|a,\lambda)\,,
\\
 P(B|a,b,A,\lambda)=P(B|b,\lambda)\,,
\end{array}
%P(A|a,b,\lambda)=P(A|a,\lambda)~ \textrm{and}~  P(B|a,b,\lambda)=P(B|b,\lambda),
\end{align}
i.e., the conditional probability of obtaining $A$ is independent of both $B$ and $b$ given the specification $\lambda$ and $a$, and analogous for the probability of obtaining $B$. Using the definition of conditional probability one trivially obtains the condition
\begin{align}
P(A,B|a,b,\lambda)=P(A|a,\lambda)\,P(B|b,\lambda)\,,
\end{align}
i.e., the joint probability for obtaining outcomes $A$ and $B$ factorizes into a product of individual probabilities for the two spatially separated systems, with each factor containing conditionalization only on local beables.  This well-known factorisation condition is thus derived from the principle of local causality, just as Bell himself stressed\footnote{``Very often such factorizability is taken as the starting point of the analysis. Here we have preferred to see it not as the \emph{formulation} of ``local causality'', but as a consequence thereof.'' Bell (1990) \cite[p.~109]{bell90}}.

%So much for Bell himself\footnote{See Norsen (2007) \cite{norsen07}  and Norsen (2009)\cite{norsen09} for an extensive and illuminating discussion of Bell's own furhter thoughts about his criterion of local causality.}. 

In the following we will subject this reasoning and mathematical formalisation to a critique. We believe that Bell's qualitative statement in the long quote above can benefit greatly from a more refined and detailed mathematical discussion than available in the literature, especially concerning (i) the understanding of the notions of sufficiency, completeness and redundancy involved;  and (ii) the fact that the settings $a,b$ and outcomes $A,B$ have very different theoretical roles in the candidate theories under study, something which, we will argue, is not adequately reflected in the reasoning leading up to  \eqref{LCmath}. The upshot of our critique is a more detailed, sharp and clean mathematical expression of local causality. It will furthermore be shown\footnote{Here only a preliminary investigation of this is worked out. A full analysis of the repercussions of our proposal is to be presented in future work, see Seevinck \& Uffink (2010, \cite{inprogress}).}  that this pays off when interpreting this condition.

\forget{
Bell considered his prelimenary qualitative formulation of local causality to be insufficiently sharp and clean for mathematics, probably because of the presence of the terms `cause' and `effect' which resist clean mathematical formalisation  (Norsen, 2007) \cite{norsen07}. His final formulation indeed does not contain these terms, however 
it does involve the notions of `sufficient', `complete' and `redundant' which, as we will show, are in need of clarification.
Furthermore, the settings $a,b$ and outcomes $A,B$ have very different theoretical roles, and we will argue that this is not adequately reflected in the reasoning leading up to  \eqref{LCmath}.
}
%%%%
\section{Cleaning up the intuitive idea}
\label{cleaning_up}
\begin{figure}[h]
\includegraphics[scale=0.6]{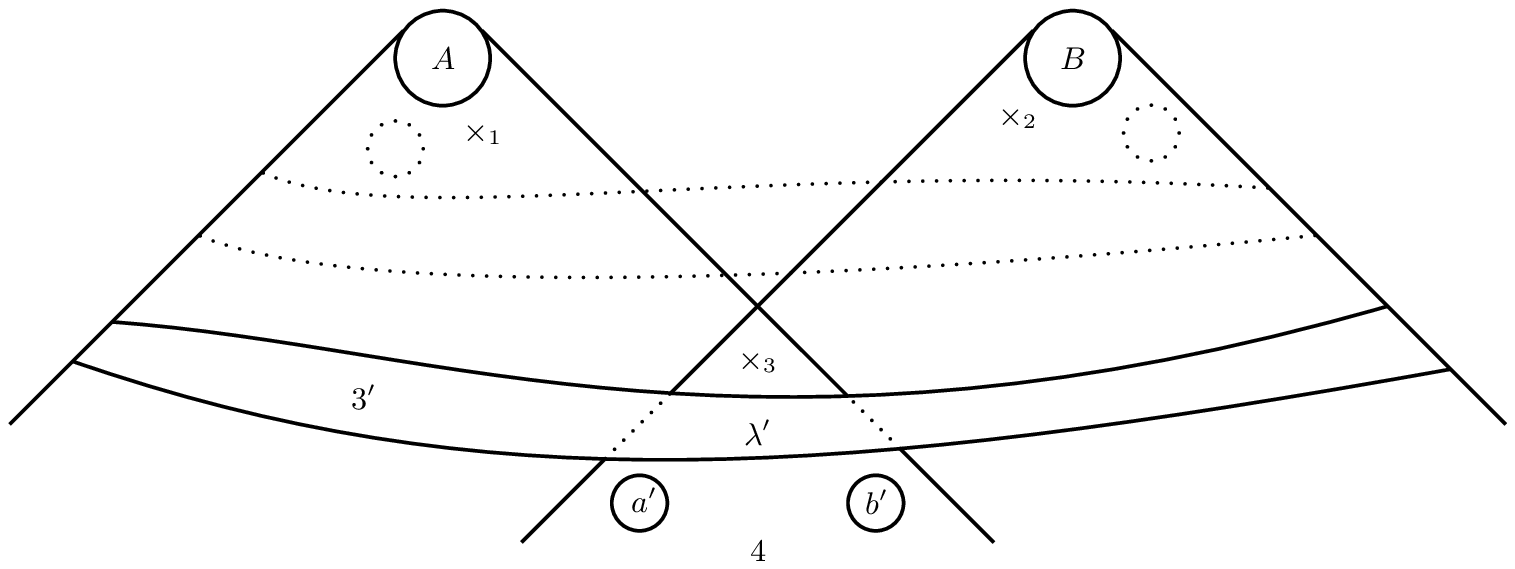}
\caption{Alternative space-time diagram of the Bell-type setup.  
%Two experimenters, often called Alice and Bob, each measure an observable $a$ and $b$, respectively, so as to %obtain outcomes $A$ and $B$. SAY MORE. 
For explanation, see text, and compare to Fig. \ref{figBell_fact}. Figure adapted from Norsen (2009) \cite{norsen09}.
} 
\label{figBell_fact_notes}
\end{figure}

\noindent
Let us first comment on some crucial aspects of Bell's formulation of local causality, most of which can already be found in the literature\footnote{\label{worthwhile}Nevertheless, most of these aspects are not well-known nor generally appreciated by commentators. It is thus worthwhile mentioning them here.}. Here we will be deliberately short because our main point lies elsewhere as will soon become clear.  When possible we will quote Bell so as to let the `master speak for himself'.  For a more detailed discusion of these points, see more of Bell himself (Bell, 1976 \cite{bell76}, 1977 \cite{bell77}, 1981 \cite{bell81}, 1990 \cite{bell90}), and especially Norsen (2007 \cite{norsen07}, 2009 \cite{norsen09}). 
%
%[ Although much more can be said, we will mainly use quotes from Bell to indicate these points, some of which will then %later be commented on. ]
%
%
\\\\
{\bf (i)~} It is important to note that the condition of local causality is only intended to be a constraint on \emph{candidate theories}, and not on the real world. %On the need for candidate theories in the quantitative statement: see Norsen 0707.0401 , blz 6
Indeed, Bell starts by writing: ``A \emph{theory} is said to be locally causal if \ldots'' [emphasis added]. Furthermore, as Norsen (2009, \cite{norsen09}) has pointed out, Bell has emphasized this point very clearly in (Bell, 1977 \cite[p.~101]{bellspeakable}):
\begin{quote}
``I would insist here on the distinction between
analyzing various physical theories, on
the one hand, and philosophising about the
unique real world on the other hand. In this
matter of causality it is a great inconvenience
that the real world is given to us once only.
We cannot know what would
have happened if something had been different.
We cannot repeat an experiment changing
just one variable; the hands of the clock
will have moved, and the moons of Jupiter.
Physical theories are more amenable in this
respect. We can \emph{calculate} the consequences
of changing free elements in a theory, be they
only initial conditions, and so can explore the
causal structure of the theory. I insist that
[local causality] is primarily an analysis of certain kinds of physical theory.''
\end{quote}
%Given a candidate theory one may thus ask whether this theory respects local causality, i.e., does \eqref{LCmath} hold? 
Note that the fundamental concepts involved such as  `beables', `completeness', and `free variables' 
%(that are to be commented on shortly) 
are all relative to some particular candidate theory. This will be become clear next.
\\\\
{\bf (ii)~} Bell uses the term `beable' to denote whatever is posited by the candidate theory to correspond to something physically real:
\begin{quote}  
``The beables of the theory are those elements which might correspond to elements of reality, to things which exist. Their existence does not depend on ÔobservationÕ. Indeed observation and observers must be made out of
beables. I use the term `beable' rather than some more committed term like `being' or `beer' to recall the essentially
tentative nature of any physical theory. Such a theory is at best a \emph{candidate} for the description of nature. Terms like `being', `beer', `existent', etc., would seem to me lacking in humility. In fact `beable' is short for maybe-able'.'' (Bell, 1984 \cite[p.~174]{bellspeakable})
\end{quote}
The concept `beable' is thus theory-relative, and it is important that the candidate theory in question is absolutely clear about what it posits as physically real (Norsen, 2007 \cite {norsen07}). Indeed, Bell emphasizes `\ldots you must identify in your theory `local \emph{be}ables'. The \emph{be}ables of the theory are those entities in it which are, at least tentatively, to be taken seriously, as corresponding to something real.'' (Bell, 1990, \cite[p.~100 ]{bell90}). And, "[i]t is in terms of local beables that we can hope to formulate some notion of local causality." (Bell, 1976, \cite[p.~53]{bellspeakable}). When applied to our particular setup of Fig. \ref{figBell_fact} this implies that the candidate theory in question must provide a well-specified account of the beables $\lambda$ in region 3.  It should be noted that $\lambda$ is not resticted in any way, as it can be anything the theory posits as physically real. In particular $\lambda$ need not be some classical hidden variable.
\\\\
{\bf (iii)~} It is important that in Fig. \ref{figBell} ``[r]egion 3 completely shields off from 1 the overlap of the backward light cones of 1 and 2.'' (Bell, 1990 \cite[p.~106]{bell90}). Likewise, in the paradigmatic setup of Fig.~\ref{figBell_fact} it is necessary that region 3a shields off region 1 from the overlap of the backward light cones of 1 and 2, and, analogously, region 3b shields off region 2 from the overlap of the backward light cones of 1 and 2 (see Fig. \ref{figBell_fact}).  Why? For if this was not the case, such as for region $3'$ in Fig. \ref{figBell_fact_notes}, then a violation of \eqref{LCmath} fails to indicate the presence of some sort of non-local causation. 
%See Norsen (2007, 2009) for a detailed explanation. 
In short, consider for example an event $\times_3$ in Fig.~\ref{figBell_fact_notes} located in the overlap of the backward light cones of regions 1 and 2 but in the forward light cone of region $3'$. Since $\times_3$ lies in the overlap of the backward light cones of regions 1 and 2, it can influence both $A$ and $B$. Now suppose $\times_3$ is a genuinely stochastic event, not predictable on the basis of the beables in region $3'$, then specification of events in region 2 could tell about $\times_3$, which in turn, could allow one to infer more about the events in 1 than is possible from just the original specification of $3'$.

The condition \eqref{LCmath} with region 3 replaced by region $3'$ --to be called called (\ref{LCmath}$'$)-- would exclude any correlation between events $b,B$ and the outcome $A$, given $a$ and $\lambda$. But a failure of this condition could be perfectly compatible with local causality. 
Thus although (\ref{LCmath}$'$) ``may validly be described as a ``no correlations'' condition for regions 1 and 2, it definitely fails as a ``no-causality'' condition.'' (Norsen, 2007, \cite[p.~12]{norsen07}).
\noindent
\\\\
{\bf (iv)~}  Bell takes it to be a trivial point that besides $\lambda$, both the settings $a,b$ and the outcomes $A,B$ are also beables\footnote{A point also made by Hans Westman, private communication.}: ``The beables must include the settings of switches and knobs on experimental equipment  [settings] $\ldots$ and the readings of intruments [outcomes].''~(Bell, 1975, \cite[p.~52]{bellspeakable}. Indeed, the settings correspond to the controllable part of some measurement apparatus and the outcomes can be taken to be manifested by the final position of some pointer (or something similar, such as a black spot on a photograph, etc.) and these are all beables as they must, in any serious candidate theory, correspond to ``something physically real'' (Bell, 1990 \cite[p. 100]{bell90}).

Thus, ``[the] ``setting'' ultimately comes down to the spatial configuration of some physically real matter, %i.e., it must be reflected somehow in the beables posited by any serious candidate theory `` 
[and] \ldots  the outcome too is just a convenient way of referring to some physically real and directly observable configuration of matter, and so [these too] will necessarily be reflected in the beables posited by any serious candidate theory.'' (Norsen, 2009 \cite[p.~5]{norsen09}).

There is, however, also a very important difference between settings and outcomes that breaks the symmetry described above.  This is a consequence of the fact that, in contradistinction to the outcomes, the settings are supposed\footnote{This requirement is crucial in deriving the so-called Bell-type inequalities that are used to proof Bell's theorem (Bell, 1964 \cite{bell64}).} to be \emph{uncorrelated} to the beables $\lambda$.
%, which we of course expect to depend on $\lambda$ in any non-trivial theory. 
The reason being that ``Now we make an important hypothesis: The variables $a$ and $b$ can be considered to be \emph{free}\footnote{Bell also calls them ``free variables -- say at  the whim of experimenters''. (Bell, 1976 \cite[p.~61]{bellspeakable})}, or \emph{random}.'' (Bell, 1990, \cite[p.~109]{bell90}). Bell continues: ``we can imagine these settings being freely chosen at the last second by two different experimental physicists, or some other random devices. If these last second choices are truely random, they are not influenced by the variables $\lambda$. Then the resultant values for $a$ and $b$ do not give any information about $\lambda$. So the probability distribution over $\lambda$ does not depend on $a$ or $b$'', i.e.,  
\begin{align}
\label{bellfree}
\rho(\lambda|a,b)=\rho(\lambda).
\end{align}
with $\rho$ some density. 

This `free variables' assumption has the important repercussion that, despite the fact that from a physical point of view  outcomes and settings are nothing but beables, they do have a completely different \emph{theoretical role} to play in the candidate theories in question.  In the literature this crucial difference has not been correctly incorporated in the mathematical formalisation of Bell's notion of local causality.  One of the main points of the present paper is that this will be performed here (in section \ref{diff_sett_out} and beyond).
\\\\
{\bf (v)~} Consider the settings $a'$ and $b'$ that are located in region 4 of Fig. \ref{figBell_fact_notes}. Since region $3'$ (just like region 3 of Fig. \ref{figBell_fact}, by the way) shields off both region 1 and 2 from the settings $a'$ and $b'$ that are located in region 4, we expect that if indeed the beables $\lambda'$ in region $3'$ are \emph{completely specified} that the beables of region 4 would be redundant for prediction of events in region 1 or 2.  The reason being that local causality is supposed to enforce that a complete specification of $\lambda'$ renders the events in the backward light cone of region $3'$ redundant for predictions of events in the future light cone of $3'$. Indeed, this is precisely what Bell's intuitive formulation claims (see the very first quote by Bell in section \ref{intuition}, and also Fig.~\ref{figcausal}). Therefore, under this understanding of local causality, and where we take $\lambda'$ to be completely specified, we expect that \eqref{LCmath} reduces to 
\begin{align}\label{expect}
\begin{array}{l}
P(A|a',b',\lambda')=P(A|\lambda')\,,
\\
 P(B|a',b',\lambda')=P(B|\lambda')\,.
\end{array} 
\end{align}

However, this ignores the fact that the settings $a'$ and $b'$ are by assumption  independent  of the beables $\lambda'$ of region $3'$ (see the remark (iv) above). Therefore, if one takes this into account, we expect the most complete specification of $\lambda'$ \emph{not} to shield off from region 1 and 2 the causal influences $a'$ and $b'$ have on these regions. 

Thus, we in fact expect \eqref{expect} to be violated, i.e., both $a'$ and $b'$ could very well (in a locally causal way) supply information about both $A$ and $B$ that is not already contained in $\lambda'$. Yet, since $a',b'$ lie in the backward light cone of both region 1 and 2 this would in no way indicate non-local causation. 

In order to exclude such spurious violations of \eqref{LCmath} it must be the case that the settings should be spacelike separated and outside the overlap of the backward light cones of regions 1 and 2 (i.e., outside region 4 in Fig. \ref{figBell_fact_notes}), and in most discussions this is indeed the case, e.g., in Fig. \ref{figBell_fact}.

Note that the main point here is not to argue for the settings to be located as in Fig. \ref{figBell_fact} rather than as in Fig. \ref{figBell_fact_notes} --for that is a rather obvious\footnote{The settings $a$,$b$ need not lie in the forward light cone of region $3$, all that is necessary is that they lie outside the overlap of the backward light cones of region 1 and 2.} if one wants to test local causality--, but that this is necessary to exclude spurious violations of \eqref{LCmath}. 
%As far as we know this point is not stressed anywhere in the literature before.
%
\\\\ 
{\bf (vi)~} The previous remark (i.e., (v)) serves as a starter for the following subsequent discussion. For that remark shows not only that settings should have a very particular space-time configuration in a sound formulation of local causality, it also indicates that the supposition that $\lambda'$ could be a complete specification of the beables in region $3'$  of Fig. \ref{figBell_fact_notes} is an \emph{illusion}; and the same holds for $\lambda$ in region 3 of Fig. \ref{figBell_fact}. This is explained next.
  
Indeed, we can not but give up the idea that $\lambda$ is a complete specification of beables in region 3, for otherwise 
``one wonders how $a$ and $b$ could possibly \emph{not} be causally influenced by $\lambda$ (in a locally causal theory).''(Norsen, 2009 \cite[p. 283]{norsen09}). In other words,
%
%since the settings $a$ and $b$ are postulated to be free variables and thus independent of $\lambda$, and given that the %backward light cones of $a$ and $b$ overlap with the region containing $\lambda$, this enforces that $\lambda$ cannot be %expected to give a complete specification of the beables in this region
%
$\lambda$ cannot be expected to be a complete specification of region 3 because one must allow for the possibility of traces in region $3$ of the causal past of both the settings, and given the independence of $\lambda$ and the settings, these traces cannot be included in $\lambda$\footnote{\label{norsen_notfree}``For example, some candidate theory (and this is actually true of every
serious extant candidate theory) might provide a specification
of the state of the particle pair which is sufficient
in the relevant sense, even though it leaves out some fact
(say, the millionth digit of the energy of some relic microwave
background photon that happens to fly into the
detection region just prior to the measurement) which
actually exists in the relevant spacetime region. Such
a fact could then be allowed to determine the setting $a$
without introducing even the slightest evidence for the
problematic sort of correlation between $a$ and $\lambda$.
%problematic sort of corelation between $a$ and $\lambda$. 
Indeed, this is just an exaggerated version of what happens
in the actual experiments, where carefully-isolated and
independent pseudo-random-number generators are used
to produce the settings at the two stations.'' (Norsen, 2009, \cite[p.~283]{norsen09})}. 

An alternative understanding of this point is that one is here faced with a dilemma. That is, the following two assumptions cannot both hold: (i) the free variables assumption, and (ii) the assumption that $\lambda$ is completely specified, i.e., contains the description of all and every beable in region 3. However, this dilemma can in fact be avoided. Because, luckily, there is no need for a completeness requirement when formalising the notion of local causality. It is only needed that the specification is \emph{sufficiently} specified, in the relevant sense. This will be futher worked out in the next subsection.

Although Bell seemed to be aware of the crucial distinction between a sufficient and complete specification of the beables involved (see next subsection), he nevertheless repeatedly\footnote{\label{foot}Most notably this is the case in the formulation of local causality as given in section \ref{intuition} above: ``And it is important that events 3 be specified completely.'' (Bell 1990 \cite[p.106]{bell90}), and ''Invoking local causality and the assumed completeness of $c$ and $\lambda$, \ldots ''(Ibid,  p. 109).
But also already in 1975 in `The theory of local beables' Bell requires this form of completeness:
``However, in the particular case that [$\lambda$] contains already a \emph{complete} specification of beables in the overlap of the two light cones, supplementary information from region 2 could reasonable be expected to be redundant.'' (Bell, 1975, \cite[p. 54]{bellspeakable}). 
And in `EPR correlations and EPW distributions', Bell writes: ``In a locally-causal theory, probabilities attached to values of local beables in one space-time region, when values are specified for \emph{all} local beables in a second space-time region fully obstructing the backward light cone of the first, are unaltered by specification of values of local beables in a third region with spacelike separation from the first two.'' (Bell, 1986, \cite[p. 200]{bellspeakable}).} stressed the need for $\lambda$ to be a \emph{complete} specification of the beables in region 3. The reason for this is the following.

Suppose the events in region $3$ are not completely specified, i.e.,  $\lambda$ leaves out some relevant beables. Then a violation of \eqref{LCmath} can no longer be used to infer some sort of non-local causation, for it could well be that the beables excluded from $\lambda$ influence both regions 1 and 2  in a locally causal way such that correlations between 1 and 2 can no longer be explained by an appeal to $\lambda$ alone. 
In order to exclude such spurious violations of \eqref{LCmath}, Bell requires the specification of region 3 by $\lambda$ to be complete\forget{because this  forbids theomission of any relevant beables}. Norsen (2007,~\cite[p. 8]{norsen07}) emphasizes that the key assumption by Bell is that events be specified \emph{completely}\footnote{But in a different paper Norsen (2009, \cite{norsen09}) mentions that, strictly speaking, such a form of completeness is not at all needed. See next subsection where this is discussed.}: "It is only because $\lambda$ is assumed to be a complete description, that the non-dependence of the probability of $A$ on the distant outcome $B$ follows from local causality.''

One final remark concerning the completeness (or sufficiency; see next subsection) that is at stake. This notion is not to be understood in the sense of the true maximal amount of knowledge concerning the systems under study. To the contrary, it is relative with respect to possible candidate theories. Therefore, the question whether the candidate theory in question is complete in the sense of including `all that really exists' is not at stake (there is no need for `omniscience', see Norsen (2007, \cite[p.~8]{norsen07}).  It thus is the candidate theory in question that should indicate when the desired completeness (or sufficiency) is achieved.
%, and any serious theory indeed does so.
\\\\
\centerline{------}\\\\
\noindent
The remarks (iv) and (vi) indicate that crucial elements of Bell's condition of local causality have escaped careful mathematical formalisation. Most notably, the theoretical distinctions between settings and outcomes 
and between the notions of sufficiency and completeness have as of yet not been properly implemented. %whereby it is the sufficiency that should be implemented and the completeness functions merely as an example of %when the sufficiency is expected to hold.

In the next two subsections these two issues are to be further worked out so as to provide the means for a clean and sharp mathematical formalisation of the notions at stake; something which is to be undertaken in section \ref{formalsuff}. This fills in the gap that was left untouched in the analysis of others, notably the careful analysis by Norsen (2007 \cite{norsen07}, 2009 \cite{norsen09}).

\subsection{On sufficiency and completeness}
\label{prelim_suff}
\noindent
The above remarks (v) and (vi) indicate that
%$\lambda'$ cannot be expected to be complete because there are traces in region $3'$  of the causal future of the %settings, and given the independence of $\lambda'$ and the settings  $a'$ and $b'$, these traces cannot be included in %$\lambda'$. DEZE ZIN BETER UITWERKEN.
%in the same vein,  
%
%likewise: 
$\lambda$ cannot be expected to be a complete specification of region 3 because one must allow for the possibility of traces in region $3$ of the causal past of the settings, and given the independence of $\lambda$ and the settings, these traces cannot be included in $\lambda$ (see footnote \ref{norsen_notfree}).

%DEZE ZIN BETER UITWERKEN. citaat shimony, Hirne and CLauser?

Bell seems to account for this by indicating, in his formulation of local causality as given in section \ref{intuition}, that the specification must be ``sufficiently specified, \emph{for example} by a full specification of local beables'' [emphasis added]. Indeed, a complete specification is not necessary, but  merely sufficient to 
interpret (1) as a condition of local causality.  However, and this is the main point, we have just seen that such a  ``complete specification''  is not an option. It is the ``sufficient specification'' that can at most be required. We must thus focus our mathematical formalisation on this notion of sufficiency.

Although both Bell\footnote{Apart from the citation on page \pageref{figBell} above (``\ldots is already sufficiently specified, for example \ldots''), taken from La Nouvelle Cuisine (Bell, 1990, \cite{bell90}), at at least two other occassions Bell mentions that the notion of sufficiency is needed when formalising the notion of local causality:
%  In 1980 in `Atomic-cascade photons, quantum-mechanical nonlocality':
``Consider, then, the hypothesis that $A$ and $B$ fluctuate independently when the relevant causal factors, at time $ T-\delta-\epsilon$ say, whatever they may be, are \emph{sufficiently well specified} \ldots .''
[...]. That is assume there are variables $\lambda$ and some probability distribution $\delta$ such that (2) holds. 
[where (2) is:  $\rho(A,B|a,b)=\int\int d\lambda \,d\mu \,\sigma(\lambda,\mu)\rho_1(A|a,\lambda)\rho_2(B,|b,\mu)$] 
(Bell 1980, \cite[p.~106]{bellspeakable}) [emphasis added]. And, %in 1981  in `Bertlmann's socks and the nature of reality':
 ``It seems reasonable to expect that if \emph{sufficiently} many such causal factors can be identified and held fixed, the \emph{residual} fluctuations will be independent, i.e., $P(M,N|a,b,\lambda)=P_1(M|a,\lambda)P_2(N|b,\lambda)$, where  [\ldots] $\lambda$ denotes any number of other variables that might be relevant. (Bell (1981), \cite[p.~152]{bellspeakable} (first emphasis added). 
 However, despite all this, Bell did not carefully distinguish between these two notions, and in fact seemed to prefer the requirement of completeness. See footnote \ref{foot} above.} (1990, \cite{bell90}) and Norsen (2009, \cite{norsen09}) indicate that completeness is not necessary, but only sufficient, in the main discussion of the notion of local causality they nevertheless gloss over this and assume that $\lambda$ provides a complete specification of  the beables in region 3. Here we want to avoid this, and therefore will proceed to give a sound analysis of the notion of sufficiency involved.
 
%However, as pointed out in remark (v) above, it is not plausible that a candidate theory would give a complete %description.

%\forget{
It might be true that ``Once one realizes that the notion of a complete specification of beables is relative to a given candidate theory, there is no further problem understanding the meaning of ``complete'' or ``full''. \ldots
% (There is only the problem of deciding which theory is true!) 
But it is less clear, even given some well defined candidate theory, what partial specifications of the beables might be considered ``sufficient''.'' (Norsen, 2007, \cite[p.~16]{norsen07}) and that  ``[i]n a more careful discussion the notion of completeness should perhaps be replaced by that of sufficient completeness for a certain accuracy, with certain epsilonics.'' (Bell, 1977, \cite[p.~104]{bellspeakable}).
But this is a practical problem that one encounters when formulating adequate candidate theories; it is not a problem for mathematically formalising the notion of  local causality. The reason being that we may assume that a serious candidate theory provides us with an unambiguous specification of the causally relevant beables.  Otherwise the question of whether the theory is locally causal does not even come up\footnote{See Norsen (2007, \cite[p.~10]{norsen07}) for a similar analysis that shows that  it should be trivial to decide in a given candidate theory what is and what is not a causal influence.  This is endorsed by Cartwright \& Jones (1991, \cite[p.~229]{cartjones}) who write "... but we take the point to be perfectly general: there are no tests of [local] causality outside of models which already have significant causal structure built in."}.

%}

Combining the above points, a first crucial observation is the following. The notion of sufficiency should be relativized with respect to a specific subclass of all beables in the candidate theory, namely the subclass of all beables in a region 3 except for the free variables and the beables causally related to the latter in that region.

%Combining the above points of remarks (i) to (vi), a first crucial observation is the following. 
%The notion of $\lambda$ being sufficient should be relativized with respect to a specific subclass, called $R_\lambda$, of all %beables in the candidate theory, namely the subclass of all beables except for the free variables and the beables causally related %to the latter.

We thus isolate a very important aspect of the notion of sufficiency involved. It is 
\begin{description}\label{alpha}
%\item for a specific purpose
\item[{$\mathrm{(\alpha)}$}] ~relative to a specific class $R_\lambda$ of beables (i.e., the beables have a particular space-time specification, and they need not include all such beables the \mbox{theory} in fact allows for!).
\end{description}
The next crucial observation comes from considering the following question: ``$\lambda$ is sufficient for \emph{what}?''  
The first obvious answer may be: sufficient for prediction of outcome $A$ or $B$. However, this cannot be true, as the probabilities assigned to region 1 need not be determined by $\lambda$ and the settings alone. Let us illustrate this with an example.  

Since we are allowing genuinely stochastic theories there could be stochastic causal relevant factors above region 3 and  within the backward light cone of 1. Consider for example some such causal factors located at events $\times_1$  or  $\times_2$ in Fig. \ref{figBell_fact_notes}. It is perfectly legitimite\footnote{See Norsen (2009, \cite[p.~12]{norsen09}) for an extensive discussion of the following point: ``The claim made in the local causality condition is not that the probabilities assigned (to events in region 1, on the basis of complete information about region 3) are the ``best possible'' probabilities the theory allows. They aren't. Better ones might be assigned, e.g., if we move region 3 forward in time, into the more recent past of region 1.''}  in a locally causal theory that these events are allowed to supplement $\lambda$ in determining the probabilities of events in regions $1$ or $2$ respectively, i.e., $\times_1$ influencing  region 1 and $\times_2$ influencing region $2$.

The correct answer to our question ``sufficient for \emph{what}?''  is that the specification $\lambda$ should be sufficient for \emph{rendering $B$ and $b$ redundant for the task of specifying the probability of outcome $A$ occurring}.
Indeed,  Bell writes:  ``The hypothesis is that any such information about 2 becomes redundant when 3 is specified completely''. As we have just seen,  `completely' should here be replaced by `sufficient' or `sufficiently complete'.

We thus isolate another very important aspect of the notion of sufficiency involved. The specification $\lambda$ should be sufficient
\begin{description}\label{beta}
\item[{$\mathrm{(\beta_1)}$}]for a specific purpose, namely  
\item[{$\mathrm{(\beta_2)}$}]to render some other variables redundant for the task of determining some particular quantity.
%\item relative to a specific class of beables
\end{description}
In the next section the notion of sufficiency as spelled out via the requirements $\alpha$, $\beta_1$ and $\beta_2$ will
be properly mathematically formalised. But before we can do so a final necessary preparatory analysis needs to be presented in the next subsection.
%WAAR MOET DIT KOMEN?:

%ad 3) the next section will show that when formalising this notion in matheatics tow distinct notions are involved: 
%functional and statistical sufficiency

\subsection[The different theoretical role of settings and outcomes]{The different theoretical role of\\ settings and outcomes}\label{diff_sett_out}
\noindent
In the mathematical formalisation above, (i) the outcomes $A,B$, (ii) the beables $\lambda$ \emph{and} (iii) the settings $a,b$ appear as  conditioning arguments in a probability distribution. See e.g. Eq. \eqref{LCmath}. 
However, if one treats the settings $a$ and $b$ as conditioning arguments in a probability distribution, this implies, at least in Kolmogorovian probability theory, that they are random variables, and thus a probability distribution over their possible values is defined within the model: one cannot write $p(x|y)$ unless $p(y)$ is also defined. In other words, this means that the candidate theory in question would have to specify how probable it is that Alice will choose one setting $a_1$ rather than $a_2$, and similarly for Bob and for their joint choises.

But that would be a remarkable feat for any physical theory. Even quantum mechanics  leaves the question what measuremtn is going to be performed on a system as one that is decided \emph{outside} the theory, and does not specify how much more probable one measuremeny is than another. It thus seems reasonable not to require from the candidate theories that they describe such probabilities. This explains, we hope, the different theoretical status of  settings $a,b$  and random variables $A, B, \lambda$. And lest one should despair that this distinction introduces a dichotomy not introduced by Bell, we note that Bell also attributed a special status to $a,b$ by relegating them the status of `free variables', meaning that their values could be set or changed at the last instant. See above, section \ref{cleaning_up}, remark (iv). This view, we believe, should also count as an argument against the presumption that a putative candidate theory ought to provide their probabilities.

However,  above (section \ref{cleaning_up}, remark (iv)) it was also mentioned that from a fundamental point of view both settings and outcomes are nothing but beables. But we have just indicated that they have very different roles in any putative candidate theory we envisage, and this means that we should not regard them on equal footing, at least theoretically. Although their ontological (or physical) status might be the same, their theoretical status is not. And this is crucial. Especially since local causality refers to putative candidate theories only (see section \ref{cleaning_up}, remark (i)). Thus, the difference between the two must be adequately reflected in any candidate theory.

Therefore we adopt the following point of view. Outcomes $A,B$, as well as the beable specification $\lambda$ are random variables and figure as arguments of a joint probability distribution $P_{a,b}(A,B,\lambda)=P_{a,b}(A,B|\lambda)\,\rho(\lambda)$. The measurement settings $a,b$ appear as labels of this probability function, not as arguments. The fundamental conditional probabilities  to be considered are thus 
\begin{align}
P_{a,b}(A,B|\lambda),
\end{align}
 instead of Bell's $P(A,B|a,b, \lambda)$. Also, Eq. \eqref{bellfree} encoding the `free variables' assumption should in fact read: $\rho_{a,b}(\lambda)=\rho(\lambda)$.

\section{Introducing mathematics:  formalizing sufficiency}
\label{formalsuff}
\noindent
Then how are we to mathematically implement Bell's idea of "$\lambda$ being sufficiently specified so as to declare redundant some of the conditional variables" in $P_{a,b}(A,B|\lambda)$, where the latter are in fact to range over both the labels $a,b$ and the random variables $A,B$?  This we will perform next. It is important to realize that two distinct notions of sufficiency are in play, i.e., where firstly the label $b$ (or $a$), and, secondly, the random variable $B$ (or $A$) becomes redundant for the task of determining the specific probability to obtain outcome $A$ (or $B$).  Each of these two notions will be clarified next.

\subsection{Functional sufficiency}
\noindent
%
%DIT NA STATISTICAL SUFFICIENCY OMDAT 'PURPOSE' EN 'RELATAVE TO' DAAR PAS AANGEGEVEN WORDEN?
%OF IS DIT AL IN EEN VAN DE VORIGE SUBSECTIES AAN BOD GEKOMEN?????????
%
%In the case of functional dependence $f_x$ where $x$ labels the different function $f_x$ the sufficiency of a variable $y%$ is rather straightforward: it means that given $y$ the function $f$ does not differ for different values of $x$, i.e., 
%\begin{align}
%f_x=f.
%\end{align}
%Local causality is to enforce that given $\lambda$ the probability distribution $P_{a,b}(A|B,\lambda)$ is sufficient for %rendering the label $b$, secondly the random variable $B$ redundant for the taks of determining this probability 
%
The first kind of sufficiency where one of the labels $a, b$ becomes redundant can be rather easily formalized. Consider a function $f_x(y)$ where $x$ labels the different functions $f_x(y)$. When variable $y$ is sufficient for the purpose of declaring $x$ redundant for task of determining the function $f_x(y)$, relative to the class of all $y$ in some specific range $R_y$,  then, given $y$ the function $f$ does not differ for different values of $x$, i.e., 
\begin{align}\label{functional}
f_x(y)=f(y),~~ \forall  y \in R_y, ~\forall x.
\end{align}
This can of course trivially be generalized to functions $f_{x_1,x_2,\ldots}(y_1,y_2, \ldots)$  that have more than one label $x_1,x_2,\ldots$, and more than one argument $y_1,y_2,\ldots$.

Recall that our preliminary analysis of sufficiency in section \ref{prelim_suff} showed that such a requirement for physical theories implies that certain variables should be sufficient \emph{for a particular purpose}  and \emph{relative to a specific class of variables}. See the requirements $\alpha$, $\beta_1$ and $\beta_2$ on page \pageref{beta}.  %KLOPT DIT, IS DIT REEDS GEDAAN?
This is retained here:  $y$ is sufficient for the purpose of making $f_x(y)$ independent of $x$ (thus not sufficient for determining its value, or for some other purpose), and this is relative to a specific class, namely to all such $y$ that lie in  a given range $R_y$.

Applying this condition to the usual Bell-type framework we have been sketching in the previous section is rather easy. First of all, we will suppose the condition \eqref{functional} to hold for each of the probability distributions $P_{a,b}(A,B|\lambda)$, $P_{a,b}(A|B,\lambda)$, etc. Secondly, $\lambda$ will play the role of $y$, and thus $R_\lambda$ the role of $R_y$, and, thirdly, the settings  $a,b$ play the role of the labels $x$.

%R_y those y do not have support on a spacetime region in overlap of the future light cone of 3 and the backward %light cone of 1, i.e., 'in between' regions 1 and 3.

Let us next turn to sufficiency in the case of statistical dependence $P( \cdot|x,y)$, and that requires considerably more clarificatory effort. 

\subsection{Sufficiency in statistical inference}
\noindent
The concept of sufficiency in the context of the theory of statistical inference was developed by R.A. Fisher (1922) \cite{fisher22}.

The basic problem of statistical inference may be formulated as follows. Suppose we have some probabilistic experiment with a fixed set of possible outcomes $x \in X$ and a family of
 probability distributions  $p_\theta$, $ \theta \in \Theta$, each of which provides some candidate description for the experiment.  Here, $\Theta$ represents some arbitrary index set.  For each value $\theta \in \Theta$,  $p_\theta(x)$ then provides the probability of $x$ to occur.
 It is assumed however that we do not know exactly what the correct probability distribution is for the experiment and the problem is to infer something about which probability distribution out of the given collection would provide a best ``fit''  for the experiment on the basis of recorded outcomes.

 It is generally useful to present the problem from a slightly expanded version, by adding the supposition that it is possible to repeat the performance of the experiment under i.i.d conditions (i.e.,  independent and identically distributed repeated trials).  In that case, assuming one performs $n$ such trials, the probability of obtaining a sequence $(x_1, \ldots, x_n)$
 is
 \beq p_\theta( x_1,\ldots, x_n) = \prod_{i=1}^n p_\theta (x_i).
  \enq
 The goal now becomes to make an inference about $\theta$ on the basis of the outcomes $(x_1, \ldots, x_n)$.

 There are many forms in which such an inference could be casted. For example, it could take the form of a point estimator, i.e., by designing a function $\tau: X^n \mapsto \Theta$ such that $\tau(x_1 ,\ldots, x_n)$ would represent  the best estimate of $\theta$.  These details need not concern us now because the concept of sufficiency is equally important in all such approaches.

 To introduce the notion of sufficiency it is useful to consider
 a set of $n$ independent functions $\{y_1, \ldots, y_n\}$ defined on $X^n$, such that the equations
 \BEQ y_1 ( x_1, \ldots, x_n) &=&c_1 \nn\\
   \vdots && \vdots \nn\\
 y_n ( x_1, \ldots, x_n) &=&c_ n \ENQ
 always have a unique solution. Thus, one might simply think of $(y_1, \ldots, y_n)$ as an alternative coordinate system that charts the points in $X^n$ just as well as $(x_1, \ldots, x_n)$. Clearly,  it is then equivalent whether one provides the recorded data  in terms of the original values $(x_1 \ldots, x_n)$ or in terms of the alternative coordinates $(y_1, \ldots, y_n)$.
 The probability distribution can be transformed to the alternative coordinates:
 \beq \label{suff1}\hat{p}_\theta (y_1 , \ldots, y_n) = p_\theta(x_1, \ldots, x_n) |\frac{\partial x_i}{\partial y_j}|  \enq
 where the last factor represents the Jacobian of the transformation.

 Now suppose that $\hat{p}_\theta$ has the following form
% \beq   \hat{p}_\theta (y_1 , \ldots, y_n) =f_\theta(y_1) g( y_1 , \ldots, y_n) \enq
\beq \label{suff2}  \hat{p}_\theta (y_1 , \ldots, y_n) =\hat{p}_\theta(y_1) g( y_1 , \ldots, y_n).
 \enq
 In that case, the function $y_1$ is said to be \emph{sufficient} for $\theta$.

 Of course, there are many choices for such an alternative coordinate system.  It is straightforward to show that if $y_1$ out of the set $\{y_1, \ldots, y_n\}$ is sufficient for $\theta$, then, the same will hold for an alternative set $\{y_1, \tilde{y_2} , \ldots, \tilde{y}_n\}$, provided that this also gives a regular coordinate system.

 The intuitive idea behind this notion of sufficiency
 is that all the information that the data provide about the unknown value of $\theta$ is in this case really contained in $y_1$ alone, because the probabilities of the values of the remaining variables $y_2, \ldots, y_n$ are insensitive to $\theta$. In other words, whatever form our inference about $\theta$ is going to be, it seems reasonable to make it depend only on the value of $y_1$,  since all the other data are irrelevant for this purpose.  Of course, if a sufficient statistic can be found this greatly simplifies the problem of statistical inference because the number of relevant data can then be reduced from $n$ to 1.  

In Fisher's own words, the criterion of sufficiency is:
\begin{quote}
``That the statistics chosen should summarize the whole relevant information supplied by teh sample. [\ldots] In mathematical language we may interpret this statement by saying that if $\theta$ is the parameter to be estimated,  $\theta_1$ a statistic which contains the whole of the information as to the value of $\theta$ which the sample supplies, and $\theta_2$ any other statistic, then the surface of distribution of pairs of values $\theta_1$ and $\theta_2$ for a given value of $\theta$ is such that for a given value of $\theta_1$, the distribution of $\theta_2$ does not involve $\theta$. In other words, when $\theta_1$ is known, knowledge of the value of $\theta_2$ throws no further light upon the value of $\theta$.'' Fisher (1922) \cite[p.~317]{fisher22}.
\end{quote}

%[GEBRUIK STREEFLAND OM DIT MEER PRECIES TE MAKEN? Leid zijn vgl. 34 af?]
It might be worthwhile to illustrate this by a simple example. Suppose we have a real-valued outcome $x$ and a collection of normal probability distributions which differ only in their location parameter, e.g.:
 \beq 
 p_\theta (x) = \frac{1}{\sqrt{2 \pi}} e^{- (x-\theta)^2/2}~.
 \enq
 It is well-known that for this case, the function
 \beq 
 y_1 = \frac{1}{n} \sum_{i=1}^n x_i 
 \enq
 provides a sufficient statistic for $\theta$.
 In general, however, the class of probability distributions for which a sufficient statistic exists is severely limited; the Pitman-Koopman  theorem implies they exist only for the exponential family.

It useful to slightly rewrite the condition of sufficiency \eqref{suff2}. Using \eqref{suff1}, we can rewrite it %\eqref{suff2} 
into $\hat{p}_\theta(y_1, y_2,\ldots, y_n)=\hat{p}_\theta(y_1)\hat{g}(y_1,\ldots, y_n)$. Next, using the definition of conditional probability this gives 
\begin{align}
\hat{p}_\theta(y_2, \ldots, y_n|y_1)=\hat{g}(y_1,\ldots, y_n),
\end{align}
which states that if $y_1$ is sufficient for $\theta$ then the conditional probability  $\hat{p}_\theta(y_2, \ldots, y_n|y_1)$ must be independent of $\theta$ (because  $\hat{g}(y_1,\ldots, y_n)$ is), and thus
\begin{align}\label{suffalternative}
\hat{p}_\theta(y_2, \ldots, y_n|y_1)=\hat{p}(y_2,\ldots,y_n|y_1).
\end{align}
This alternative formulation of sufficiency shows that once  $y_1$ is given, the rest of the data (i.e., $\{y_2, \ldots, y_n\}$) is irrelevant to $\theta$.

For our purposes the following notes are of crucial importance. Note firstly that notions of locality or causation can be kept safely on the bench in this approach. For example, it might be that $\theta$ labels various races of tomato plants, and $x$ the weight of a tomato produced by a such a plant.

Secondly, Fisher's talk about `information' should be understood in the following sense. It refers to information that is `contained' in a sample and that is `about' something. To say that $y_1$ is sufficient is qualified by saying that it is sufficient for a purpose (inferring the value of $\theta$, it may well be insufficient for other purposes!) and relative to a class $R_y$, namely all other statistics of the same sample space, i.e., all other functions of $y_2,\ldots, y_n$ of the outcome space $X^n$, i.e., $R_y=\{y_1, \ldots,y_n\}$.

So we see that this framework naturally incorporates the two aspects mentioned in subsection 
\ref{prelim_suff}, namely that sufficiency of $\lambda$ should be characterised as (i) for a specific purpose, the purpose beign to render some other variable redundant and (ii) relevant to a specific class of variables.  See the requirements $\alpha$, $\beta_1$ and $\beta_2$ on page \pageref{beta}.
   
Finally, some words about its status. In mathematical statistics, one might feel, sufficiency is actually just a name for a particular definition. As such, one may ask, how can this be of any help to foundational or conceptual problems in physics? However, even in statistics sufficiency is much more than just a definition. It is a desideratum. But it can be turned into something much more powerful by formulating the demand that, for certain physical probabilistic or stochastic theories, certain variables should be sufficient (in the sense given by the definition above) for a particular purpose relative a specific class of variables. To this we will now turn.
 
\subsubsection{Bayesian inference}\noindent
The above approach used the orthodox formulation of statistical inference, in which parameters are kept strictly distinct from outcomes.  The basic reason for this division is that while a statistical model provides probabilities for the events or outcomes $x$, there is usually not a corresponding probability for parameters, i.e., they are usually not random variables  but labels for the probability distributions indicating settings of measurement apparata.

At first sight this dichotomy between parameters and outcomes in orthodox statistical inference corresponds nicely to the same dichotomy adopted above in section \ref{diff_sett_out} between the theoretical roles played by the settings $a,b$ and outcomes $A,B$?  However, in the context of Bell's notion of local causality we are not aiming at making an inference about the settings $a$ or $b$. Therefore we will have to change the perspective somewhat. 

There is an alternative approach to statistical inference, the Baysian approach, that does not rely on such a strict division between events and parameters. 

  Bayesian statistical inference proceeds from a similar point of view except that now one assumes the existence of a so-called prior probability distribution over  the parameter $\theta$.  Furthermore, the probability distributions
  $p_{\theta}(x)$ are now reinterpreted as conditional distributions 
  \beq p_\theta(x) = p(x | \theta)~.
  \enq
Given these two assumptions,  it is possible to provide a so-called posterior probability distribution by means of Bayes' theorem, i.e.,
  \beq p(\theta | x ) = \frac{ p(x|\theta) \rho (\theta)}{\int d \theta p(x|\theta) \rho(\theta)}~. \enq

Extending this to the case of multiple, independent and identically distributed trials one obtains:
  \beq p(\theta | x_1 , \ldots x_n)  = \frac{ \prod_{i=1}^n  p_\theta(x_i) \rho(\theta)} { \int   \prod_{i=1}^n  p_\theta(x_i) \rho(\theta) d\theta}~. \enq

In the  Bayesian approach, the goal of  statistical inference is
to report this posterior probability distribution. Note that this viewpoint does not necessarily presuppose a subjective interpretation of probability.

As stated before, sufficiency is an important notion regardless of
which precise approach to statistical inference is chosen. In the Baysian approach this takes the following form.
Firstly, note that the transformation \eqref{suff1} remains valid. Secondly, 
consider $\hat{p}(\theta | y_1,\ldots,y_n)$ and note that using Bayes Theorem this is equal to 
\begin{align}
\frac{\hat{p}(y_1,\ldots,y_n|\theta)\rho(\theta)}{\int \hat{p}(y_1,\ldots,y_n|\theta)\rho(\theta)d\theta}~.
\end{align}
Then assuming $y_1$ to be sufficient for $\theta$, i.e.,  assuming \eqref{suffalternative} obtains, we deduce that 
\begin{align}
\hat{p}(\theta | y_1,\ldots,y_n)
&=
\frac{\hat{p}_\theta(y_1) \hat{g}(y_1,\ldots,y_n)\rho(\theta)}{\int \hat{p}_\theta(y_1) \hat{g}(y_1,\ldots,y_n)\rho(\theta)d\theta}
\nn\\&
=
\frac{\hat{p}_\theta(y_1)\rho(\theta)}{\int \hat{p}_\theta(y_1)\rho(\theta)d\theta}~.
\end{align}
Thus once the fundamental assumptions of Bayesian inference are in place, the sufficiency condition is
entirely equivalent to
 \beq    \hat{p}(\theta  | y_1 , \ldots y_n) = \hat{p} (\theta |  y_1) 
 \label{suffeq}\enq
 Perhaps, this makes the underlying motivation of  sufficiency even clearer: if the above condition holds, then the probability of $\theta$, once $y_1$ is given, is not changed when the values of $y_2, \ldots, y_n$ are included. These additional functions of the data are irrelevant or redundant for the purpose of assigning the posterior probability. In accordance with the above terminology, this is expressed as: $y_1$ is sufficient for the purpose of rendering some set of other variables redundant relative to a class of variables $R_y=\{y_1,\ldots,y_n\}$.

It should also now be clear how this condition would fit in with the usual Bell-type framework we have been sketching in the section \ref{intuition} (e.g., see Fig. \ref{figBell_fact}). First of all, we will suppose the condition \eqref{suffeq} to hold for each of the probability
distributions labeled by the settings $a$, $b$.  Secondly, $\lambda$ will play the role of $y_1$, and each of the $y_i$ ($i\neq 1$) should be one of the other random variables in the candidate theory in question, for example one of the settings $A,B$, or some other beable specification $\lambda'$. The role $R_y$  is thus taken over by $R_\lambda$. Thirdly, $\theta$ is one of the outcomes $A,B$.

 Now, although obviously $\lambda$ is not a datum nor a function of the data, it is still assumed to be a random variable. It has a value, and although it will in general be hidden for us we can reason about the hypothetical case that we would know its value. Perhaps one might wonder if the identification $\theta$ with the outcomes $A,B$ is a valid move to make. %WAT BEDOELT JOS HIER?
After all, we have introduced the variables $y_1,\ldots,y_n$ as functions on the space of outcomes $X^n$, whereas $\theta$ labelled probability distributions. But in a Baysian approach this distinction does not count as fundamental anymore. As is clear from \eqref{suffeq}  we are basically dealing with probability distributions on $\Theta \times X^n$, and both $\theta$ and $y_i$ can be seen as functions on this larger space.

By now we have gathered enough mathematically clean and precise tools to explicate Bells' intuitive notion of local causality, and this will be carefully outlined in the next section.

\section{The baby: Bell's local causality `mathematically sharp and clean'}\label{babysection}
\noindent
Recall that our analysis of sufficiency showed that such a requirement for physical theories implies that certain variables should be sufficient (in one of the two senses given above) for a particular purpose relative to a specific class of variables. We have argued that then \eqref{functional} and \eqref{suffeq} can be obtained as mathematical criteria when dealing with, respectively, labels attached to functions and random variables that are to be conditioned on.

In Bell's conditionn of local causality it is the beable specification $\lambda$ that is supposed to be sufficient. Thus in order to analyse this condition as a sufficiency criterion, we must, firstly, indicate the purpose for which $\lambda$ is sufficient (per requirement $\beta_1$ and $\beta_2$) and, secondly, indicate relative to which beable class this is so, i.e., to indicate the range $R_\lambda$ of allowable $\lambda$ (per requirement $\alpha$).

We believe that by now it should come as no surprise that our proposal is the following: 
Consider again Fig. \ref{figBell}. A candidate theory will be said to be locally causal when, if region 1 is space-like separated from region 2, the theory provides a specification $\lambda$ in a region 3 that shields off region 2 from the overlap of the back-ward light cones of 1 and 2, that is sufficient, in the sense of \eqref{functional} and \eqref{suffeq}, for the purpose of rendering the beables in the far-away region 2 redundant for the task of predicting the probabilities of events in region 1, where $\lambda$ is relative to the class of all allowable beable specifications that can be given about a region 3 (according to the candidate theory and consistent with the `free variables' assumption), i.e., $R_\lambda$ contains all allowable beable specifications in this region 3.%, but that does not contain any free variables related to the settings.

If we now apply this to the standard bi-partite setup of Fig. \ref{figBell_fact} we obtain that if the candidate theory in question obeys local causality then the theory provides a beables specification $\lambda$ in region 3a that is sufficient for the purpose of rendering the far-away outcome $B$ and setting $b$ redundant for the task of determining the probabilities of obtaining $A$, and this is so relative to all other allowable beable specifications $R_\lambda$ the candidate theory in question provides for region 3a. Using the results of the previous section this entails that functional sufficiency \eqref{functional} renders the label $b$ and statistical sufficiency \eqref{suffeq} the random variable $B$ redundant for prediction of $P_{a,b}(A|B,\lambda)$. For determining the probability of obtaining outcome $B$ a completely analogous analysis obtains so that we finally get the mathematically sharp and clean formulation of the condition of local causality:
\begin{align}
\begin{array}{l}
P_{a,b}(A|B,\lambda)=P_a(A|\lambda),\\
P_{a,b}(B|A,\lambda)=P_b(B|\lambda),
\end{array}\label{LCcorrect}
\end{align}
from which one trivially obtains factorisability:
\begin{align}\label{FACTcorrect}
P_{a,b}(A,B|\lambda)=P_a(A|\lambda)\,P_b(B|\lambda).
\end{align}
Although rather similar to \eqref{LCmath}, which in section \ref{intuition} was claimed to be the mathematical expression of the condition local causality, the differences and alternative derivation are crucial. 
%This will be commented on next.

%\subsection{Remarks}
\section{Not throwing out the baby with the bathwater}\label{notout}
%further relevance}
\subsection{Remarks}
\noindent
{\bf (one)~} All this might look like an overcomplicated way to obtain the already well-known, i.e., something similar to \eqref{LCmath}, but we have in fact obtained quite a lot:  a mathematically clean formulation  \eqref{LCcorrect} that brings to the forefront crucial aspects otherwise left out. The formalisation, firstly,  encodes the particular notions of sufficiency and redundancy that are involved, secondly, incorporates the theoretical distinction between outcomes (random variables) and settings (labels) enforced by the `free variables' assumption, and finally,  indicates rigourously where the constraints set by the notions of locality and causality enter the mathematical formalisation.

This latter point has not been stressed before and needs some elaboration. 
Note that the mathematical formalisation of sufficiency itself needs no requirement what so ever of locality or causality, as can be deduced from the formal analysis of section \ref{formalsuff}. We must thus look elsewhere, and here is our proposal. When applying both statistical and functional sufficiency to the setup of Fig.~\ref{figBell_fact}, requirements of locality and causation necessarily come in play when fullfilling the requirements $(\alpha)$ and $(\beta_1)$ and $(\beta_2)$ of section \ref{prelim_suff}, page \pageref{alpha}.

Firstly, it is by an appeal to the principle that causality can only be local in the sense of Fig.~\ref{figcausal}, that the purpose for which $\lambda$ is sufficient is specified. Indeed, because local causality stipulates that causes operating in a certain region in space-time must lie in the backward light cone of that region and effects of anything occuring in that region can \emph{only} lie in its forward light cone, we get the inference that anything outside the backward and forward light cone of that particular space-time region should be causally redundant. See Fig. \ref{figcausal} and the very first quote by Bell in section \ref{intuition}. But in order to distinguish mere correlation from causal influence, it is, secondly, that the range $R_\lambda$ of allowable beable specifications is restricted by considerations regarding locality and causality. The beables in $R_\lambda$ must lie in a spacetime region with specific characteristics, namely it must shield off the overlap of the light cones of region 1 and 2 from these same regions 1 and 2. 
The possibility of spurious violations of local causality mentioned in remark (iii) on page \pageref{figBell_fact_notes} is in this way eliminated.
 
%Where does the locality do any job: not in the factorisation, but in the specification of $\lambda$!
%The factorisation is motivated by the sufficiency of $\lambda$ 
%Importantly, causation and locality issues are involved. in specifying the purpose, and in specifying the class of beables.

In conclusion, local causality is not a mere statistical no-correlations requirement; to the contrary,  it has a substantial relation to the relativistic causal structure of Fig. \ref{figBell}. See also Norsen (2007, \cite{norsen07}) who extensively argues for this point. According to our analysis presented here, Bell's condition of local causality is a special form of statistical and functional sufficiency, where the purpose for which $\lambda$ is claimed to be  sufficient, and the range of validity (the class $R_\lambda$ of beable specifications to which it is relative) are motivated by locality and causality constraints.

It is important to realize that the notions of locality and causality referred to here involve nothing but the special relativistic causal structure as exemplified in the light cone structure of Fig. \ref{figcausal}. Importantly, there is no need for some philosophical theory of causation or an appeal to the problematic notion of Reichenbach's Principle of the Common Cause, or the like (see also Norsen (2007, \cite{norsen07})). 
%
%{\bf (viii)~} No need for an appeal to the problematic notion of Reichenbach's Principle of the Common Cause. 
%
%WIL IK DE OPMERKINGEN VAN JOS GEBRUIKEN, DIE OOK OVER EBF GAAN, DAN MEOT IK JOS DAAROM VRAGEN, WANT %IK HEB DIE NEIT IN ZIJN LATEX FILE.
%
%HOUD HET KORT: eignelijk wil je alleen kwijt dat er geen appeal aan reichenbachian notiond of common cause gedaan hoeft te worden.
%
%0707.0401 Blz 10, Norsen: against Butterfield en Brown [26,27]. Ik ben het daarmee eens we laten zien dat het principle of %Reichenbachian common cause niet nodig is.   (ook against Chris Timpson).
%Compare to other analyses of the underlying motivation, appealing to the principle of common causes, Stochastic Einstein %Locality, etc.  (as Uffink does) Copy the criticism by JOS from, his paper.
%
%Elby, Brown, Foster: Jarrett-completeness is motivated by statistical completeness, and by considerations of locality and causality.  %(But they do so by appeal to Reichenbach's common cause principle). We think this is bad. Also there is no spacetime description.  %This is unfortunate.
%Hoe zit het met het partitoneren van de ruitme $lambda$ en Elby Foster BRown's idee van splitting up the ensembles in %subensembles? Dit is een vorm van volledigheid.  van het vergelijken van verschillende theorieen met deeper HV's.
%
\\\\
{\bf (two)~}\label{(two)} It is only $\lambda$ that is supposed to be sufficient, and not $\lambda$ \emph{plus} one or both of the settings. Of course the settings $a$ and $b$ are needed to determine the outcomes $A$ and $B$ respectively, but, remember, prediction of outcomes of measurement is not the purpose for which $\lambda $ is supposed to be sufficient!

As was mentioned in footnote~\ref{norsen_fout}, Norsen (2007 \cite{norsen07}, 2009 \cite{norsen09}) does include the local setting in the specification of the beables that are supposed to render some other beables in a space-like separated region redundant.  However, this is not needed, and in fact even unwanted\footnote{Also, it should be mentioned that Bell himself explicitly refrains from doing this. See Fig. (6.6) of Bell (1990 \cite[p.~108]{bell90}).}.  
%
% because if specification of a local setting is indeed required for this purpose, then there would be correlations between the causal %pasts of the local setting and the space-like separated region. 
%
For, after all, what should be in $R_\lambda$? Only those beables whose causal past could be correlated, in the sense of Fig.~\ref{figcausal} and according to the candidate theory in question, to the causal past of the beables that are to be rendered redundant, i.e., the beables in region $2$, such as $B,b$.  Being free variables the local settings should thus not be in $R_\lambda$, as the following example shows. 

If including $a$ in the beable specification of region 3 could be relevant to render $b,B$ in region 2 redundant for prediction of the probability to obtain $A$,  then there needs to be a genuine possibility  in the candidate theory under study for correlations between the causal past of $a$ and the events in region 2 (see Fig.~\ref{figBell} and  Fig.~\ref{figBell_fact}). However, this is excluded by the `free variables' assumption. We can think of $a$ being chosen at the very last instant, and therefore (in any locally causal theory)  $a$ can make no difference as to whether beables in space-like separated region 2 are, or are not redundant for prediction of obtaining outcome $A$ in region 1. Furthermore, from a mathematical point of view, including $a$  in $R_\lambda$ makes no difference and is therefore better left out.
\forget{ For being a free variable, $a$ is not expected to influence (in a locally causal theory) the outcome of measurement in region $2$. And neither is the causal past of each of the choise of settings expected to be relevantly correlated  to measurement on the far away system.
      Thus for the purpose of rendering $b$ redundant we need only include $\lambda$ there is no need to also include $a$. if including $a$ would make a difference there is a correletion between outcome $B$ and the choise of measurement $a$ or anything in its causal past. This is precisley what local causality intends to exclude.
KLOPT DIT? }			
\\\\
{\bf (three)~}  One could proceed in a two step procedure to obtain the mathematical formulation of local causality \eqref{LCcorrect} and the condition of factorisability \eqref{FACTcorrect}, by firstly requiring statistical sufficiency \eqref{functional} so as to obtain, 
\begin{align}\label{two1}
\begin{array}{l}
P_{a,b}(A|B,\lambda)=P_{a,b}(A|\lambda)\,,
\\
P_{a,b}(B|A,\lambda)=P_{a,b}(B|\lambda)\,,
\end{array}
\end{align}
and only then functional sufficiency \eqref{suffeq} to subsequently get:
\begin{align}\label{two2}
\begin{array}{l}
P_{a,b}(A|\lambda)=P_a(A|\lambda)\,,\\
P_{a,b}(B|\lambda)=P_b(B|\lambda)\,.
\end{array}
\end{align}
These two requirements together indeed imply \eqref{LCcorrect} and \eqref{FACTcorrect}.

It might be tempting to think of each of these two conditions \eqref{two1} and \eqref{two2} as implementing a different weaker assumption than local causality itself.
For after all it is the conjunction of the two that gives the desired condition \eqref{LCcorrect}. Logically this is indeed true. But
despite this theoretical difference, the physical status of the two conditions is exactly the same. Both are a consequence of local causality, and the appeal to notions of locality and causality used in implementing the functional and statistical sufficiency are just the same, see remark (one) above.
% GAAT DIT NOG GEBEUREN?
Elsewhere this will be further argued for (Seevinck \& Uffink, 2010, \cite{inprogress}) and the comparison to a similar famous two-step procedure by Jarrett (1984, \cite{Jarr}) and Shimony (1984, \cite{Shim}) will be there presented.
\\\\
{\bf (four)~}
Orthodox quantum mechanics violates \eqref{FACTcorrect}. Indeed, as is well-known, the quantum mechanical predictions using the singlet state can be easily used to provide such a violation. A closer look reveals that the theory violates statistical sufficiency because it violates \eqref{two1}, but it obeys \eqref{two2} and thereby functional sufficiency.

From this we can conclude that quantum mechanics does not provide a beable specification $\lambda$  in region $3$ with the correct characteristics, i.e., the theory is unable to provide a specification of beables in any appropriate region 3 such that the outcome $B$ always becomes redundant for the probability of determining outcome $A$.  

It is tempting to draw more grand conclusions than this one, say, of a somewhat foundational, philosophical or meta-physical nature. We will refrain from doing that here, but in remark (six) below we will address the controversial question of what legitimate conclusions can be drawn from violations of local causality.
\\\\
{\bf (five)~} The qualification of the class of beables $R_\lambda$ is essential because we must carefully ensure that 
no spurious violations of local causality are allowed for, and at the same time that we do not encode too much in the specification $\lambda$ so as to come into conflict with the `free variables' assumption.  Therefore, as was shown earlier, the specification cannot be taken to be complete, as some free variables must probably be left out, yet it must be allowed to range over all other beable specifications the candidate theory in question allows for. This justifies calling the specification of  $\lambda$  `sufficiently complete', meaning that nothing causally relevant that is not `free' from the candidate theory under consideration is left out, without claiming the specification to be complete in the sense that everything in the theory is included, including the free variables. This ensures that the spurious violations of remark (iii) and (iv) can not occur, and that local causality is not just a mere no-correlations condition. 
%
%Instead of requiring a complete specification, we require 'sufficiently well-specified'  relative to the class of ALL possible HV %descriptions of this region that the candidate theory allows for. So in that sense it can not be completed, since all are considered. %Hoort dit bij subsectie  'two notions of sufficiency'?
\\\\
{\bf (six)}  Suppose \eqref{LCcorrect} fails, how are we to interpret this? Well, given our remarks above, there seems to be only one option, namely that local causality is violated, i.e., there must be some non-local causation present in the candidate theory under study.

We cannot blaim a violation of \eqref{LCcorrect} on the specification $\lambda$ not being sufficiently well-specified, for as was just argued in remark (five) above, $R_\lambda$ ranges over all allowable beable descriptions and should be regarded `sufficiently complete'.

Nor can we blaim such a violation on the existence of `locally explicable' correlations. The space-time structure of Fig. \ref{figBell_fact} and the further specification of local causality via the notion of sufficiency, including the requirements ($\alpha$), ($\beta_1$) and ($\beta_2$), excludes any spurious violations of \eqref{LCcorrect} due to correlations that do allow a locally causal explanation.   
This we argued for in section \ref{cleaning_up}.
%If this was not the case the notion of 
% (in de subsectie: two notions of sufficiency? refer to Elby Brown Foster?  only intuitively compelling when... 

The question then arises what it (philosophically, meta-physically) means for local causality to be violated.  Such an investigation will be performed elsewhere \cite{inprogress}.

%, but a preliminary investigation is presented in the next, final,  subsection.
%\\\\
%
%{\bf (six)} Then:  if local causality fails, then either: DIT HAD IK JUIST EERDER VERWIJDERD. zie verwijderd.tex
%\\\\
% MAAR DAT DOE IK OOK HIERONDER AL!

%\forget{
%{\bf (seven)~}
As a final remark we wish to present a point made by Norsen (2009, \cite[p.~12]{norsen09})\footnote{See footnote \ref{worthwhile} for the reason of mentioning it here.} concerning a violation of \eqref{LCcorrect}:
%``Bell has carefully set things up so that a violation of
%[\eqref{LCcorrect}] entails that there is some non-local causation.
``It isn't necessarily that something in region 2 is causally
influencing something in region 1, or vice versa. It is
always possible that there is some other event, neither in
region 1 nor region 2, which was not determined by [$\lambda$],
and which itself causally influences both [beables in region 1] and [in region 2]. The
point is, though, that this causal influence would have
to be non-local (i.e., would have to violate the special relativistic causal structure sketched in [Fig. \ref{figBell}].''

\section{Envoi}  %Discussion
\label{envoi}
\noindent

Have we thrown out the baby with the bathwater? We believe not, as our cleaning up of the intuitive idea of local causality for mathematics has proven to be fruitful and to clarify hitherto unknown aspects of the notion of local causality. It should be seen as a complement to Norsen's manuscripts (2007 \cite{norsen07}, 2009 \cite{norsen09}) taking his analysis a step further.

Also, we believe that our mathematical formalisation of Bell's notion of local causality further unearths its rich conceptual background, and that it thereby brings us  a bit closer to answering the hard and open foundational questions that arise from attempts to incorporate violations of local causality into our physical worldview.  %But, of course, we are a long way from doing that. 
 
What critical light this novel mathematical formalisation throws on other interpretation and motivations of Bell's notion of local causality is still to be worked out. This we hope to unearth in the near future (Seevinck \& Uffink, 2010, \cite{inprogress}).

\begin{acknowledgements}
\noindent
MPS acknowledges very fruitful conversations with Hans Westman and Eric Cavalcanti and thanks the Centre for Time, Sydney, Australia for hosting him as a guest researcher. JU acknowledges fruitful discussions with Joe Henson and thanks the Perimeter Institute, Waterloo, Canada for generous hospitality.
\end{acknowledgements}

%\appendix*
%\section{een}
%\section{twee}

\end{document}